\documentclass[aps,pra,reprint,superscriptaddress]{revtex4-2}
\usepackage{graphicx,amsmath}
\usepackage{xcolor}
\usepackage[english]{babel}
\usepackage{gensymb}

\graphicspath{{PaperFigs/}}

\begin{document}
\title{Noise Resilience in a High-Bandwidth Atom Interferometer}
\author{Jonathan M. Kwolek}
\email{jonathan.m.kwolek.civ@us.navy.mil}
\affiliation{U.S. Naval Research Lab, 4555 Overlook Ave SW., Washington, DC 20375}
\author{Sunil Upadhyay}
\affiliation{Amentum, 4800 Westfields Blvd, Suite \#400, Chantilly, VA 20151}
\author{Adam T. Black}
\affiliation{U.S. Naval Research Lab, 4555 Overlook Ave SW., Washington, DC 20375}

\begin{abstract}
The utility of inertial sensors depends on resilience against real-world dynamics and noise. Atom interferometry offers a sensing technology with the advantage of good long-term stability, high sensitivity, and accuracy. High measurement bandwidth improves an atom interferometer's ability to reject errors due to dynamics and noise. Here we demonstrate resilience against time-varying environmental noise by rapidly switching the direction of inertial sensitivity in the atom interferometer through a common technique known as k-reversal. We demonstrate sub-interrogation-time k-reversal at 592~Hz in a cold-beam atomic interferometer with an inverse interrogation time of 148~Hz. The interferometer fringe output is read out continuously and post-processed using nonlinear Kalman filters to determine both the inertial and error contributions to the output phase. The resulting power spectral densities show a significant reduction of phase error due to a noisy magnetic field as the k-reversal frequency increases.
\end{abstract}

\maketitle

\section{Introduction}

Inertial navigation uses acceleration and rotation rate sensors to determine a change in position over time. Both short-term sensitivity and long-term stability are important parameters in the function of inertial navigation sensors, as they contribute to error growth over time. Atom interferometry has long been a candidate technology for improved inertial sensors \cite{gustavson_precision_1997, fang_metrology_2016, jekeli_navigation_2005}, due to the propensity of atomic systems towards long-term stability, high sensitivity, and accuracy \cite{narducci_advances_2022}. Recently, demonstrations of atomic interferometers on moving platforms have been pursued to assess the viability of these sensors for applications such as inertial navigation and mobile gravimetry \cite{narducci_advances_2022, bidel_airborne_2023, kohler_at-sea_2024}. These demonstrations can help to quantify the effects of platform dynamics, sensor bandwidth, and environmental noise in sensor architectures. An important step along this path is the optimization of the control of atom-light interactions \cite{saywell_enhancing_2023} to improve sensitivity, stability, and noise immunity.

For inertial sensors used in strapdown inertial navigation applications, high measurement bandwidth is important for ultimate navigation performance due to the influence of single-axis and multi-axis undersampling or aliasing effects \cite{titterton_strapdown_2004}. For motion along a single axis only, a sensor with limited sampling rate in the presence of high-frequency vibrations may alias motion, resulting in a low-frequency error in sensor output leading to long-term navigation errors. In the presence of simultaneous multi-axis motion, aliasing errors known as coning and sculling also create biases if sensor sampling occurs at an insufficient rate. Aliasing of non-white noise sources, in addition to aliasing of inertial signals, likewise drives sensor requirements towards high operating bandwidth.

Aliasing and dynamic errors particularly present a challenge in inertial navigation applications of pulsed cold-atom interferometer sensors that operate at low measurement rate \cite{titterton_strapdown_2004, narducci_advances_2022, peters_high-precision_2001, cheinet_measurement_2008}. Comparatively high measurement rate is achievable in atom interferometers based on laser-cooled ensembles featuring recapture \cite{Mcguinness_high_2012}, in warm vapors \cite{Biedermann_atom_2017}, or in continuous atomic beams \cite{gustavson_precision_1997, kwolek_continuous_2022}. Ideally, continuous or zero-dead-time measurement can limit undersampling-induced errors that can occur in pulsed atom interferometers \cite{dutta_continuous_2016,kasevich_kinematic_2008,joyet_theoretical_2012, devenoges_improvement_2012}.

In this work, we study continuous, inertially sensitive, cold-atom interferometer measurements with calculated $1/e$ measurement bandwidth exceeding 50 Hz for both accelerations and rotations. The interferometer operates in the spatial domain, in which a continuous beam of atoms in freefall passes through a series of platform-fixed laser beams that drive coherent, momentum-changing Raman transitions in the atoms. To increase fringe contrast, signal-to-noise ratio, and dynamic response, we make use of a 3D-sub-Doppler-cooled rubidium beam, which we have described elsewhere \cite{kwolek_three-dimensional_2020}. We have previously characterized the fringe contrast and intrinsic noise performance of an atom interferometer based on stimulated Raman transitions in this cold-atom beam \cite{kwolek_continuous_2022}. Recently, there have also been a number of related demonstrations of spatial-domain interferometers employing transversely cooled atom beams \cite{sato_closed-loop_2024, meng_closed-loop_2024}.

Resilience to technical and environmental noise is a key consideration for atom-based sensors, particularly for sensors that are intended to operate outside of a controlled laboratory environment. Because the atoms are manipulated and detected using resonant optical interactions, laser intensity and frequency fluctuations can induce noise in the interferometer phase. For example, differential ac Stark shifts during Raman interactions or due to scattered cooling or detection light during free-space propagation can cause unwanted phase shifts. Meanwhile, magnetic fields can cause sensor errors due to linear or quadratic Zeeman shifts. Each of these error drivers can occur intrinsically (e.g., due to noise in current sources), due to environmental variations, or even (hypothetically) due to a deliberate jamming attack by an adversary. Shielding is able to reduce much of the impact of external magnetic fields, but adds size, weight and cost. Intrinsic error sources, on the other hand, may not be amenable to shielding in all cases. Magnetic field sensitivity of a variety of atom interferometer configurations has been studied in the past \cite{Ren_characterizing_2023, Zhou_precisely_2010, Hu_analysis_2017, Barrett_atom-interferometric_2011}. 

Typical three-pulse atom interferometer configurations \cite{kasevich_kinematic_2008, gustavson_precision_1997, kasevich_measurement_1992, peters_measurement_1999} carry a degree of intrinsic noise insensitivity for a number of reasons. These sensors are commonly operated on a two-photon Raman clock transition that is insensitive, to first order, to magnetic field. Additionally, a central $\pi$ pulse induces a spin-echo-like effect that eliminates sensitivity to clock transition frequency shifts that are temporally symmetric about the central $\pi$ pulse. Finally, sensors such as gyroscopes \cite{gustavson_precision_1997, xu_limits_2024, sato_closed-loop_2024, savoie_interleaved_2018, meng_closed-loop_2024} and gravity gradiometers \cite{mcguirk_sensitive_2002, biedermann_testing_2015, sorrentino_sensitivity_2014} that employ a differential measurement of multiple atom interferometers benefit from rejection of common-mode noise. As a result of these effects, many common light-pulse atom interferometer sensor configurations are somewhat resilient against perturbations that vary slowly both temporally and spatially. However, spatial or temporal gradients in fields causing frequency shifts of the clock transition can cause interferometer phase shifts that constitute an error source in inertial measurement.

Suppression of interferometer phase error is often accomplished through the technique of ``area reversal'' or ``k-reversal'' \cite{mcguirk_sensitive_2002,durfee_long-term_2006, Louchet-Chauvet_influence_2011}. In this technique, the direction of photon recoil imparted by stimulated Raman transitions, and hence the direction of inertial sensitivity, of the interferometer is periodically reversed. Successive measurements with opposing directions of the Raman wavevector $\vec{k}_\mathrm{eff}$ are combined to produce an inertial measurement with reduced error. In numerous prior k-reversal demonstrations, the primary concern has been the removal of static or slowly-varying error sources \cite{weiss_precision_1994,mcguirk_sensitive_2002,stockton_absolute_2011,savoie_interleaved_2018,yankelev_atom_2020}. In this work, we instead evaluate the response of a k-reversed atom interferometer to error sources that vary rapidly compared with the sensor's measurement bandwidth. We observe that, if the k-reversal frequency is not sufficiently high compared with the frequency range of both the noise and the inertial signals, k-reversal can add noise in the atom interferometer's measurement band and actually be detrimental to long-term stability of the interferometer. This is essentially due to the fact that the k-reversed outputs are measured sequentially, resulting in aliasing errors when k-reversal occurs too slowly \cite{black_time-domain_2024}.

 A significant advantage of the continuous nature of the cold atomic beam interferometer is that it enables rapid k-reversal within the time $T$ separating interferometer atom-light interactions, known as the interrogation time. We demonstrate k-reversal in a continuous cold-atomic beam interferometer at a rate up to 592 Hz, much faster than $1/T=148~\mathrm{Hz}$. We study the trade-space of k-reversal in the continuous-beam regime in the presence of a simulated jamming attack in the form of deliberately applied magnetic field noise. In the experiment, the noise due to the applied magnetic field has amplitude sufficient to overwhelm the true inertial signals observed in a laboratory environment across most of the inertially sensitive spectral range. We demonstrate that, by sufficiently rapid k-reversal, this applied magnetic field noise can be attenuated to a level significantly below the level of the observed inertial signal.

\section{Apparatus}
\label{sec:apparatus}
\begin{figure}
\includegraphics[width=\linewidth]{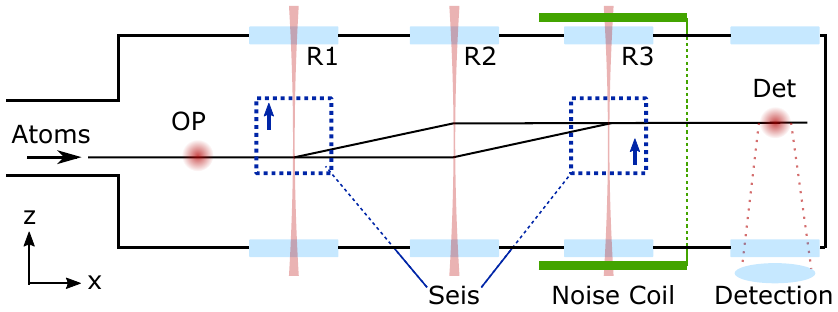}
\caption{\label{fig:apparatus} Atoms travel through the vacuum chamber along the direction of the x-axis from left to right, encountering beams for state-preparation (OP), coherent state manipulation for generating the interferometer (R1, R2, R3), and detection (Det). Wilcoxon seismometers (Seis) are mounted below the intersection points of R1 and R3 with the atomic beam. Direction of acceleration sensitivity is shown with a blue arrow. Noise coils (green) are mounted coaxially along R3. Light is relayed to a detector outside of the vacuum chamber (Detection).
}
\end{figure}

The experimental apparatus discussed here was developed and presented in previous works \cite{kwolek_three-dimensional_2020,kwolek_continuous_2022}. For clarity, we include a brief overview of the experimental setup as well as novel implementations relevant to this manuscript.

The cold-atom source consists of a two-dimensional ${}^{87}$Rb magneto-optical trap (2D MOT) and tilted three-dimensional optical molasses (3D OM), both of which operate continuously. Atoms are cooled to $15~\mu\text{K}$ in three dimensions and travel at a mean velocity of $10.8\pm0.2\;\text{m}/\text{s}$. The atomic beam passes through a graphite baffle to mitigate scattered light and absorb background atoms. In the interferometry chamber shown in Fig.~\ref{fig:apparatus}, atoms travel through a state-preparation beam that optically pumps the atoms into the $F=1, m_F=0$ state, followed by three equally-spaced Raman beams driving the $F=1,m_F=0~\leftrightarrow~F=2, m_F=0$ transition. Finally, the atomic beam passes through a detection laser beam tuned to the D2 cycling transition to induce state-selective fluorescence. An imaging system and avalanche photodiode detect fluorescence from the atomic detection region. Coils surround both the 3D OM and primary experimental chambers in order to impose a zero-field and constant linear field respectively.

The Raman transitions are driven by three laser beams that are phase-modulated using fiber-coupled ixBlue electro-optic modulators (EOMs). The optical setup of these lasers is detailed in \cite{kwolek_continuous_2022}. The three Raman beams are retroreflected from a set of three parallel mirrors glued to a common silicon carbide substrate. The Raman beams propagate at a $1\degree$ angle from normal to the atom beam, resulting in a projection of atomic longitudinal velocity along the Raman beam axis and a nonzero mean Doppler shift of $\pm 500~\mathrm{kHz}$.  As a result, either direction of Raman photon recoil can be chosen by altering the phase modulation frequency to match either sign of Doppler shift. The EOMs are driven at the ground hyperfine spacing of ${}^{87}$Rb, $6.834\;\text{GHz}$, modified by the Doppler shift. This modulation frequency is generated using a $6.734\;\text{GHz}$ tone from a National Instruments QuickSyn synthesizer combined in a single-sideband mixer with a secondary, variable ${\sim}100$ MHz signal. The $100$~MHz signal is generated using an agile Wieserlabs FlexDDS synthesizer, which enables fast modulation and switching of the resultant combined 6.8~GHz signal. Opposite signs of $k_\mathrm{eff}$ were driven using two distinct phase-coherent rf outputs on the FlexDDS sent through an rf switch. Future implementation of this technique could instead use a single agile rf source per EOM drive.

Interferometer phase is measured by adding a frequency offset $f_\mathrm{mod}$ to the EOM drive frequency of the third Raman beam. The phase of this frequency offset signal is then compared with the atomic fluorescence readout to determine interferometer phase. While this phase can be measured using a lock-in amplifier, in the present experiment we determine interferometer phase by processing the atomic fringe data using a nonlinear Kalman filter \cite{cheiney_navigation-compatible_2018}. This method has advantages in dynamic response over lock-in amplification, and is further discussed in Sec.~\ref{Sec:kswitchProcess}.

For tests in which we add magnetic noise to the interferometer, we make use of an additional ``noise coil" pair shown in Fig.~\ref{fig:apparatus}. This coaxial pair of coils is orientated along the axis of the third Raman beam, and has an internal diameter of 7 cm, wrapped around 2.75" CF flanges on either side of the atom beam. The coils are driven in series by sending white noise through a voltage preamplifier, used to apply a low-pass filter to the noise with a cutoff at 300 Hz. This choice of cutoff frequency ensures that the majority of the noise power is around the frequency range of inertial sensitivity of the atom interferometer. The signal is then fed into the control input of a ColdQuanta coil driver, which drives the current through the coils. Through measurement of the magnetic field noise inside the coils, we have confirmed that the coils produce a relatively flat noise spectrum across the 300 Hz region of interest.

\section{k-Reversal}\label{Sec:kswitching}

\subsection{Model}

The principle of noise reduction by k-reversal takes advantage of the assumption that the atom interferometer phase may be written as 
\begin{equation}
\phi_\mathrm{tot} = \kappa \phi_i + \phi_\mathrm{b} + \phi_L
\end{equation}
where $\phi_i$ is the inertial phase, $\kappa=\pm1$ is the sign of the effective Raman wavevector $k_\mathrm{eff}$ , $\phi_b$ is the bias phase due to level shifts of the clock transition, and $\phi_L$ is the phase due to the temporal phase differences of the fields driving the Raman transitions. Here, $\phi_i$ includes not only inertial phase due to rigid-body accelerations and rotations of the sensor, but also inertial error contributions due to differential path length variations and wavefront error in the Raman retroreflection paths. The latter errors are not suppressed by k-reversal, and we focus in this work only on suppression of $\phi_b$. $\phi_L$ is controlled by a direct digital synthesizer locked to a stable frequency reference and therefore may be known accurately. Control of $\phi_L$ makes possible lock-in detection of phase through modulation of the interferometer population signal. In applications designed to measure inertial effects, $\phi_b$ is an undesirable error term, typically caused by differential light shifts, quadratic Zeeman shifts, or atomic collisions \cite{peters_high-precision_2001}. 

We assume periodic k-reversal such that $\kappa(t)$ is described by a square wave with amplitude 1 and frequency $f_k$. Taking an average $\left<\phi_{tot} - \phi_L\right> = \left<\kappa \phi_i\right> + \left<\phi_b\right>$ of atom interferometer phase measurements obtained over a time $2t_k = 1/f_k$ makes it possible to estimate $\phi_b$ and $\phi_i$ individually. In the case of continuous measurement, we define the average phase over a single k-reversal period as $\left<\phi\right>\!(t)=\frac{1}{2 t_k}\int_{t-t_k}^{t+ t_k} \phi(t') dt'$. The resulting estimates of bias and inertial phase are $\tilde{\phi_b} = \left<\phi_{tot} - \phi_L\right>$ and $\tilde{\phi_i} = \kappa (\phi_{tot} - \phi_L - \tilde{\phi_b})$  respectively. These estimates are accurate in the limiting case of static $\phi_i$ and $\phi_b$, such that $\left<\kappa \phi_i\right>=0$ and $\left<\phi_b\right>=\phi_b$. Time-varying $\phi_L$ is not problematic so long as $\phi_L$ is assumed to be known at all times. 

\begin{figure}
\includegraphics[width=\linewidth]{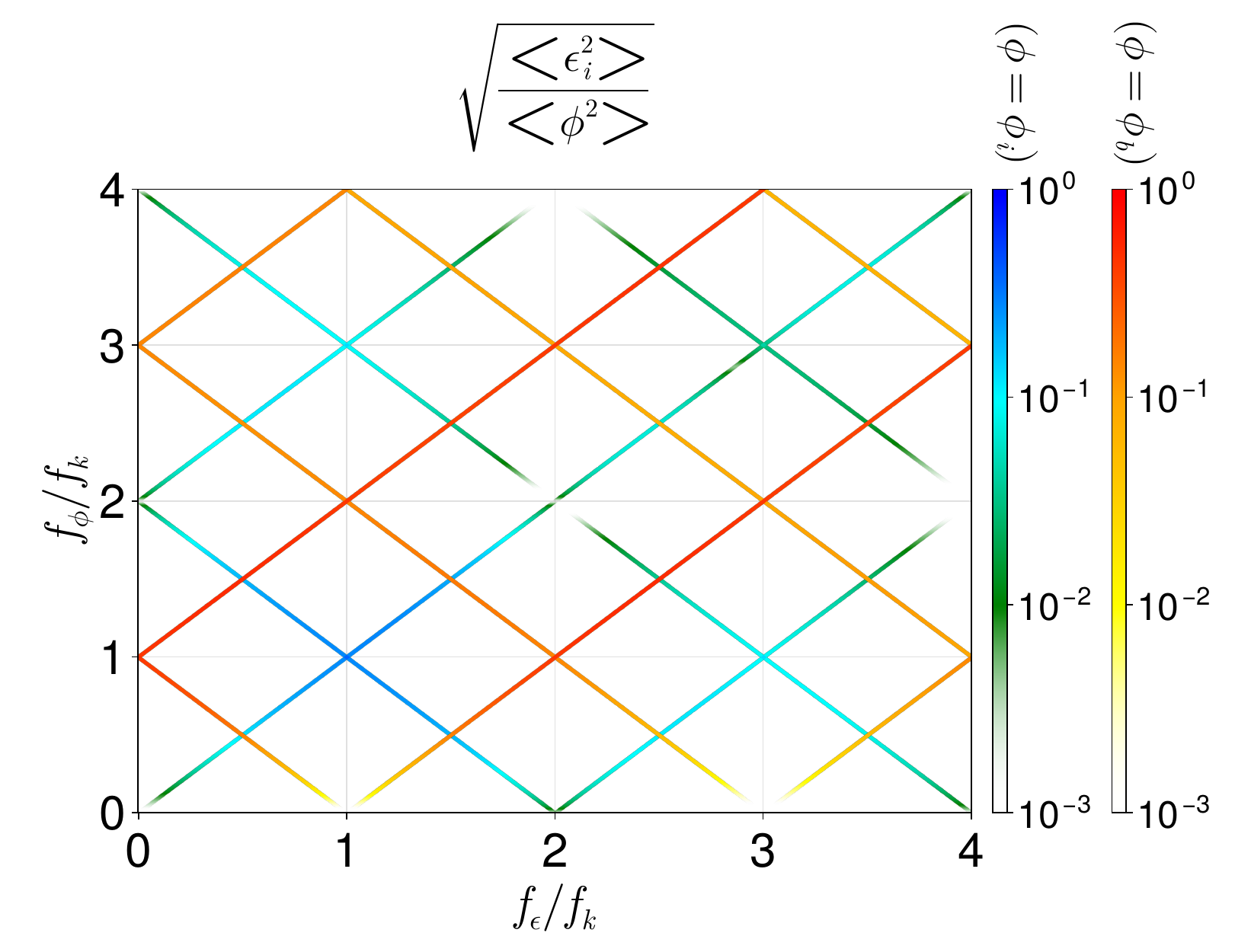}
\caption{\label{fig:alias} Calculated positive-frequency Fourier decomposition of the inertial phase error $\epsilon_i(t)$ in inertial phase estimation given by Eq.~\ref{eq:errorformula}, in the presence of k-reversal at frequency $f_k$, as a function of analysis frequency $f_\epsilon$ (horizontal axis), in response to a sinusoidally varying input phase at frequency $f_\phi$ (vertical axis). Two different limiting cases are plotted with different color scales on the same graph. Green-blue scale: input phase is due to inertial effects only, $\phi = \phi_i$. Yellow-red scale: input phase is due to bias only, $\phi = \phi_b$. 
}
\end{figure}

If $\phi_i$ and $\phi_b$ are not static, a time-varying inertial phase measurement error $\epsilon_i(t) = \phi_i(t) -\tilde{\phi}_i(t)$ is introduced. This error is purely due to the k-reversal process, and is given by  
\begin{equation}
\label{eq:errorformula}
\epsilon_i(t) = \kappa (\phi_b - \left<\phi_b\right>) - \kappa \left<\kappa \phi_i\right>.
\end{equation}
For example, time-varying bias phase $\phi_b$ at the k-reversal frequency $f_k$ results in a dc inertial phase error error $\epsilon_i$, while a component of $\phi_i$ at frequency $f_k$ results in a frequency component of $\epsilon_i$ at the same frequency $f_k$. This result has simple interpretations: a bias phase at frequency $f_k$ is misinterpreted by this algorithm as a dc $\phi_i$, while an inertial signal at frequency $f_k$ is misinterpreted as a dc $\phi_b$. As a result, the responses of both the bias estimate and the inertial phase estimate at frequency $f_k$ are strongly attenuated, as we observe below. The frequency-dependent rms magnitude of the aliasing error $\epsilon_i$ in response to sinusoidally-varying $\phi_i$ or $\phi_b$ is shown in Fig. \ref{fig:alias}. In this figure, the mean-square error is averaged over all possible relative phases of the inertial or bias signal relative to the k-reversal square wave. The figure is similar to a folding diagram used to illustrate an aliased frequency response \cite{dunn_measurement_2017}; however, in Fig.~\ref{fig:alias} the spectrum of inertial phase \emph{error} is plotted as a function of the true signal frequencies. This analysis is particular to the case of continuous measurement, as it does not include the effect of periodic sampling of sensor output that can reduce measurement bandwidth and introduce additional aliasing effects dependent on the relationships between the sampling rate, frequencies of inertial and error signals, and $T$.

The errors depicted in Fig.~\ref{fig:alias} occur due to aliasing as a result of the finite-rate sampling of the bias phase by the k-reversal method and the boxcar averaging of total phase to estimate bias phase. To avoid aliasing of temporally varying components of $\phi_i$ and $\phi_b$, a high k-reversal rate, much larger than the frequencies of inertial and bias phase variations, is desired.

\subsection{Technique}

\begin{figure*}
\includegraphics[width=\linewidth]{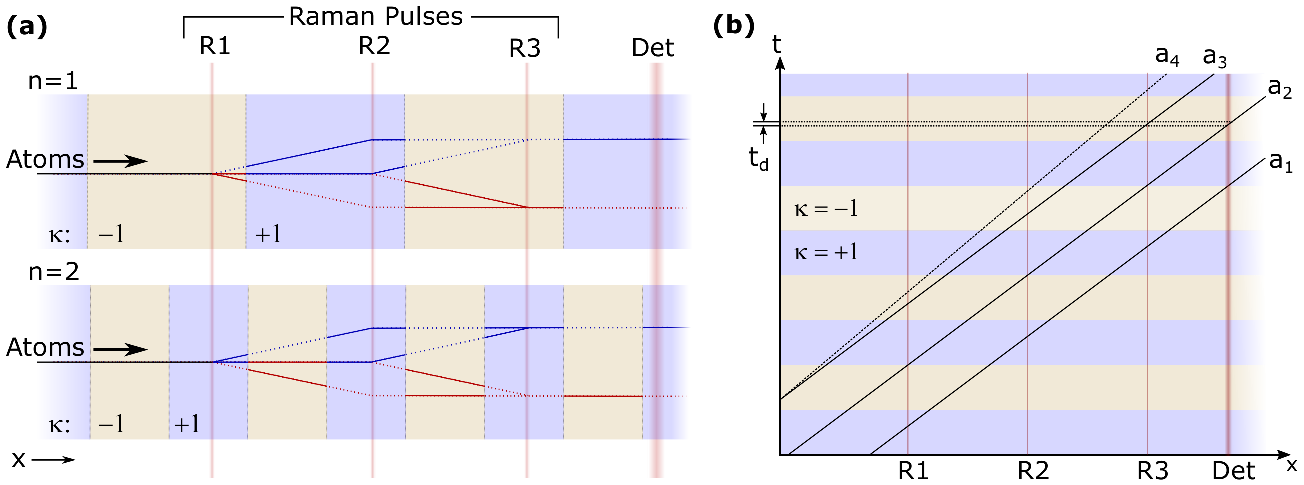}
\caption{\label{fig:kswitch}(a) Illustrations of a snapshot in time showing the atomic beam propagating through the interferometer region during k-reversal. In time, atoms traverse the interferometry region in the direction of the x-axis from left to right through the three Raman beams (R1, R2, R3) and detection region (Det). The background in the snapshot is color-coded beige ($\kappa=-1$) or blue ($\kappa=+1$) based on the sign of $\vec k_\mathrm{eff}$ atoms in each region will interact with from each Raman beam. Solid lines indicate atoms currently on a trajectory, and dotted lines indicate the parts of the trajectory without atoms. k-reversal rates $f_k = n/2T$ are shown, with $n=1$ (top) and $n=2$ (bottom). (b) Spacetime diagram of the atomic paths through the interferometer for $n=2$. Shading indicates the states of R1, R2, and R3 as a function of time, either addressing the $+\vec k_\mathrm{eff}$ or $-\vec k_\mathrm{eff}$ interferometers as indicated with blue and beige shading respectively. The width of the detection beam is a contributor to the dead time $t_d$ between different interferometer cases. Four example atomic paths are shown (diagonal lines): an atom traveling through the $+\vec k_\mathrm{eff}$ interferometer ($\mathrm{a_1}$), an atom crossing the Raman beams during the dead time between reversals ($\mathrm{a_2}$), an atom traveling through the $-\vec k_\mathrm{eff}$ interferometer ($\mathrm{a_3}$), and an atom at a different velocity for which the interferometer does not close ($\mathrm{a_4}$).
}
\end{figure*}

Our ideal implementation of k-reversal involves rapidly reversing the direction of $\vec k_\mathrm{eff}$ while maintaining a maximized measurement duty cycle. In the present experiment, k-reversal is implemented by switching the phase modulation frequency of the Raman beams as described in Sec.~\ref{sec:apparatus}. To maximize the duty cycle, the timing of k-reversal is chosen such that the interferometer closes for the largest possible fraction of the atomic position and velocity distribution. By using the appropriate sign of $\vec k_\mathrm{eff}$ in each of the interferometer beams, k-reversal can be accomplished by switching between cases every $t_k=T/n$, or at a frequency $f_k=n/2T$ for integers $n$. Examples of timing sequences for $n=1$ and $n=2$ are shown in Fig.~\ref{fig:kswitch}. In order to ensure interferometer closure, switching at frequencies corresponding to odd $n$ requires the Raman k-vectors to be oriented at any given time with alternating wavevectors for the first, second and third Raman beams. Explicitly, we define the set $K$ of $\kappa_i$ indicating the $\kappa$ value for the $i$th Raman beam. In the odd-$n$ case, $K_o=\{\pm 1,\mp 1,\pm 1\}$. Conversely, for even $n$ the Raman k-vectors must be all oriented in the same direction, such that $K_e=\{\pm 1,\pm 1,\pm 1\}$.

One drawback in a realistic implementation of k-reversal is the ``dead time" $t_d$ required for the interferometer to make the transition between cases, during which the interferometer contrast drops due to mixed contributions of the two cases to the output. For convenience of calculation, here we define the dead time as the time during which the interferometer output contains at least a 10\% contribution from both cases. The minimum achievable value of $t_d$ is set by the transit time of atoms through each Raman beam. Assuming our Raman beams have a $40\;\mu\text{m}$ waist and the longitudinal velocity of the atoms is $11\;\text{m}/\text{s}$, this minimum dead time is $t_d=4.6\;\mu\text{s}$. However, atom temperature creates a more stringent limit. Given our measured temperature of $15~\mu\text{K}$ in the direction of atomic beam propagation, a point-like atomic position distribution at the first Raman beam spreads to a $500\;\mu\text{m}$ rms width when it crosses the third Raman beam. This results in a $t_d\approx 120\;\mu\text{s}$ dead time from atomic temperature alone at the end of the interferometer. If we assume the atoms are detected at a time $T$ after the last interferometer beam, the temperature-limited dead time becomes $t_d\approx170\;\mu\text{s}$. Finally, the non-zero detection width contributes to dead time, since atoms at different positions in the detection beam are simultaneously detected. We investigated multiple detection beam sizes, minimizing switching time while also maintaining the SNR of the interferometer to ensure that the fluorescence detection was not the dominant noise source in the measurement.  The detection beam waist is $0.78\;\text{mm}$, resulting in a contribution to k-reversal dead time of $90\;\mu\text{s}$. To account for these sources of dead time, the data-processing filter implements a $200\;\mu\text{s}$ dead time at each instance of k-reversal, as discussed below.

\subsection{Signal Processing}\label{Sec:kswitchProcess}

Modulated fringe output from spatial-domain matter-wave interferometers is commonly processed using demodulation techniques such as lock-in amplification \cite{gustavson_precision_1997}. In these techniques, the sinusoidally varying photodiode signal and an rf reference signal, both at frequency $f_\mathrm{mod}$, provide the input and reference signal for a lock-in amplifier. The cosine and sine phase quadrature outputs of the amplifier are used to determine the interferometer fringe phase and amplitude. The lock-in amplifier's low-pass filter at cutoff frequency $f_\mathrm{LP} \ll f_\mathrm{mod}$  determines the measurement bandwidth of the interferometer readout. The phase output can be used to operate in a closed-loop mode of operation to improve linearity and dynamic range \cite{sato_closed-loop_2024, meng_closed-loop_2024}, but in the present experiment we operate without feedback.

\begin{figure}
\includegraphics[width=\linewidth]{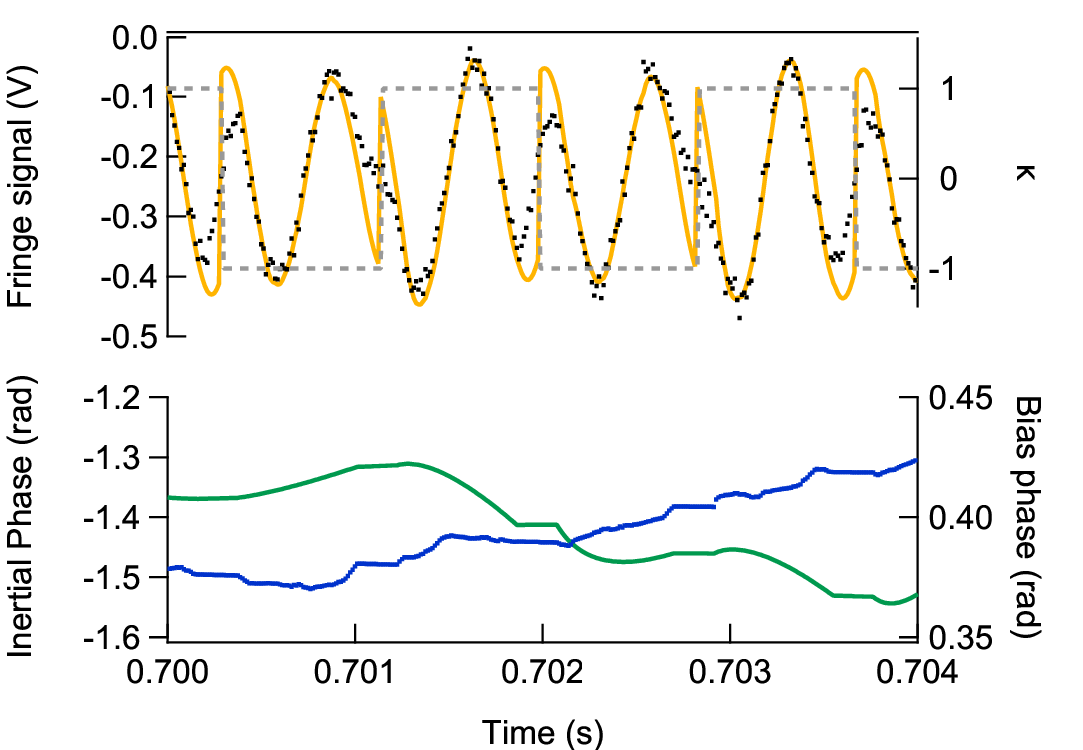}
\caption{\label{fig:kswitchreal}Example of inertial and bias phase estimation at $f_{k}=592~\mathrm{Hz}$ and $f_\mathrm{mod}=1776~\mathrm{Hz}$. Upper plot: Amplified photodiode data (black dots) measures atomic fluorescence at the interferometer output and is used to determine interferometer phase. k-reversal is indicated by the plot of $\kappa$ (grey dashed line). The filtered interferometer fringe (orange line) is based on the inertial and bias phase estimates. Lower plot: Inertial phase (blue dots) and bias phase (green line) estimated using UKF. The flat regions in phase estimation that occur during the $t_d=200~\mathrm{\mu s}$ transition time around each k-reversal edge is due to the elimination of UKF measurement updates during those periods.
}
\end{figure}

For the cold-beam atom interferometer studied here, the rapid k-reversal mode of operation creates a challenge for conventional lock-in demodulation due to the periodic phase discontinuities of magnitude $2 \phi_i$ created by k-reversal. The sign of $k_\mathrm{eff}$, denoted $\kappa$, alternates between $\kappa=1$ and $\kappa=-1$ with frequency $f_k$. The interferometer phase is then determined during each constant-$\kappa$ dwell time $t_k-t_d$. As described above, high values of $f_k$ are desirable in order to accurately measure and distinguish high-frequency inertial signals and noise. However, as described in the previous section, the maximum attainable value of $f_k$ is limited due to the finite dead time, $t_d = 200~\mathrm{\mu s}$ in the present experiment. The finite size of the detection beam also creates a low-pass filter in atomic population measurements, causing contrast reduction at high values of $f_\mathrm{mod}$. We choose to operate with $f_\mathrm{mod} = 1776~\mathrm{Hz}$ in these studies. To maintain a high measurement duty cycle $1-t_{d}/t_k >75\%$, we operate with $f_k \leq592~\mathrm{Hz}$. At the highest rate of k-reversal, a single dwell time at constant $\kappa$ contains only $f_\mathrm{mod} (t_k-t_d)=1.1$ full cycles of population modulation as shown in Fig.\ref{fig:kswitchreal}.  Under these conditions, the requirement $f_{LP} \ll f_\mathrm{mod}$ in conventional lock-in phase estimation cannot be met because the interferometer phase will change due to k-reversal before the lock-in output has settled. 

To avoid the challenges inherent in lock-in detection for estimation of time-dependent phase, we employ phase estimation based on nonlinear Kalman filters (KF) \cite{cheiney_navigation-compatible_2018,khodaparast_review_2022,dedes_error_2024}. Both $\phi_b$ and $\phi_i$ are expected to be continuous functions of time even in the presence of k-reversal, and so a KF that separately tracks each of these components, rather than the total phase alone, does not need to contend with large phase discontinuities. Compared with linear KF, nonlinear estimation produces a more accurate phase estimation (at the expense of higher computational complexity) because the fringe observation is a nonlinear function of interferometer phase. We obtained similar phase estimation results from an extended Kalman filter (EKF) and an unscented Kalman filter (UKF), ultimately choosing to use UKF. We implemented UKF using the GaussianFilters.jl package in Julia \cite{gfgithub}. To obtain interferometer fringe data, we acquire atomic fluorescence signal on a photodiode and log the photodiode data on an oscilloscope for 100~s at a sampling rate of 80~kS/s. We simultaneously log the rf local oscillator output, the digital signal controlling the rf switch that implements k-reversal, and the outputs from classical seismometers placed on the atom interferometer platform.

In post-processing, we perform a three-step analysis to estimate the inertial and bias phase. First, we use a UKF to estimate the phase of the local oscillator relative to the oscilloscope timebase as a function of time. Second, under the assumption that the bias phase does not change during a singe k-reversal dwell time $t_k$, we estimate the mean difference between total interferometer phase and the local oscillator phase during each such interval by $\chi^2$ minimization over the phase of a trial sine wave compared with the fringe data. The bias phase is pre-estimated as the windowed moving average of this estimated total phase with a window duration $2 t_k$ in order to impose the constraint that the bias changes slowly relative to $f_k$. (This bias pre-estimation step may not be strictly necessary, but we find that it improves stability of the UKF.) Finally, given the bias phase estimate and fringe data, we estimate the inertial phase at the full sampling rate using a second UKF. In this filter, observations are provided by the fringe data and the bias pre-estimate. The filter states are the fringe amplitude, offset, inertial phase, bias phase, and first derivative of inertial and bias phase. The values of pre-estimated local-oscillator phase and $\kappa$ provide control inputs. During the transition time $t_d$ between $\kappa = \pm 1$, the UKF estimation step is carried out but the measurement update step is omitted. In the present experiment, we employ modulation only via a linear phase ramp, but the UKF phase readout technique should be equally applicable to other phase modulation schemes including those that might be challenging to analyze using conventional lock-in techniques \cite{kawasaki_analyzing_2025}.

Using the UKF phase estimation technique, it is possible to track an estimated inertial phase of the interferometer at the highest update rate available. However, as discussed in Sec.~\ref{sec:seismometer} and demonstrated in Eqs.~\ref{eq:rotscalefactor}-\ref{eq:accscalefactor}, the interferometer phase response to rigid-body accelerations and rotations for frequencies comparable to or greater than $1/T$ exhibits low-pass filter behavior. Therefore, for frequencies much greater than $1/T$ the inertial phase estimate does not reflect true platform acceleration and rotation. Rather, at such high frequencies the inertial phase estimate is likely to be dominated by optical path fluctuations such as those caused by air currents and mirror vibrations, technical noise sources such as laser noise and magnetic field (as discussed in Sec.~\ref{sec:magnetic}), and quantum projection noise. In the presence of high-frequency technical phase noise sources, it is advantageous to estimate the inertial and bias phase at high rate to avoid undetected interferometer phase wraps and additional aliasing. In the present study, we low-pass filter the estimated inertial and bias phase at 1~kHz. In phase spectra examined below, we consider only frequencies below 200~Hz.

\section{Seismometer Phase Estimation}
\label{sec:seismometer}
Because the purpose of this study is to evaluate the response of the atom interferometer's estimate of inertial phase to non-inertial noise, it is valuable to have an estimate of the true inertial environment for comparison. For this purpose, we employ a pair of Wilcoxon 731A seismometers mounted to the underside of the optical breadboard housing the atom interferometer, as close as possible to the first and third Raman beam retroreflection locations, and aligned with their axes of sensitivity roughly parallel to the Raman beam propagation direction. These ac-coupled seismometers have a response flatness of 10\% between 0.1 Hz and 300 Hz and a noise spectral density ranging between $4~\mathrm{ng/Hz^{1/2}}$ and $30~\mathrm{ng/Hz^{1/2}}$ \cite{Wilcoxon}, generally below the estimated intrinsic acceleration noise of the atom interferometer \cite{kwolek_continuous_2022}. To eliminate a slow dc-coupled output drift of these seismometers from the calculation, we high-pass filter the outputs at 2~Hz. 

While the atom interferometer is sensitive to a combination of accelerations and rotations of the sensor platform, the seismometers individually are sensitive to local accelerations only. However, under the assumption that the sensor platform is nearly static and undergoes small-amplitude, rigid-body motion such that the small-angle approximation is always valid, the pair of seismometers is able to account approximately for rotational motion about the vertical axis of rotational sensitivity of the atom interferometer. We do not expect the seismometer phase estimates to precisely reproduce the phase output of the atom interferometer for various reasons including errors in our knowledge of seismometer position and measurement axis alignment, and deviations from the assumption of rigid-body motion. 

For the purpose of comparison, we use the classical seismometers to estimate the inertial component of atom interferometer phase. We double-integrate the accelerometer output to obtain the change in each accelerometer position along the Raman beam axis. We interpolate between the accelerometer positions to obtain the time-dependent position of each of the three Raman retroreflection points $z_A$, $z_B$, and $z_C$ along the axis parallel to the Raman beam propagation direction. Finally, we estimate the inertial interferometer phase as follows:
\begin{equation}
\label{eq:phiseis}
\tilde{\phi_i}^{(seis)}(t) = k_\mathrm{eff} (z_A(t-T) - 2 z_B(t) + z_C(t+T))
\end{equation}
where we choose $t$ to be the time when the atoms cross the central Raman beam. In practice, an appropriate time offset is added to account for the time of flight from the central Raman beam to the detection beam.

Due to the assumption of rigid-body motion of the Raman retroreflection mirror and Raman optical paths in the seismometer-based phase estimation, the atom interferometer sensitivity function filters the acceleration and rotation power spectra \cite{cheinet_measurement_2008}. The frequency-dependent scale factors for small-amplitude, rigid-body rotations and accelerations can be written

\begin{align}
\label{eq:rotscalefactor}
\left|\frac{\phi_i^{(rot)}}{\Omega_0}\right| &= \frac{2 k_\mathrm{eff} v T}{\omega} \left|\sin{(\omega T)}\right| \\
\label{eq:accscalefactor}
\left|\frac{\phi_i^{(acc)}}{a_0 }\right| &= \frac{2 k_\mathrm{eff}}{\omega^2}(1-\cos(\omega T)),
\end{align}
where $\Omega_0$, $a_0$ are the rotation rate and acceleration amplitudes respectively, and $\omega$ is the rotation or acceleration frequency. When $\omega \ll 1/T$, these formulas for frequency-dependent scale factor reduce to the standard formulas for dc rotation and acceleration scale factors of the three-pulse atom interferometer \cite{dubetsky_atom_2006}. From the small-amplitude formulas above, the frequency $\omega/2\pi$ for which the response is reduced by a factor of $e$ relative to the dc scale factor is approximately $0.35/T$ for rotations and $0.52/T$ for accelerations. In the present experiment, the platform vibrations are predominantly accelerational in nature at the interferometer location, as demonstrated by a strong positive correlation in the output signals of the two seismometers. As a result, the spectrum of the atom interferometer phase estimated using the pair of seismometers exhibits strong minima at frequency multiples of $1/T$.

\begin{figure}
\includegraphics[width=\linewidth]{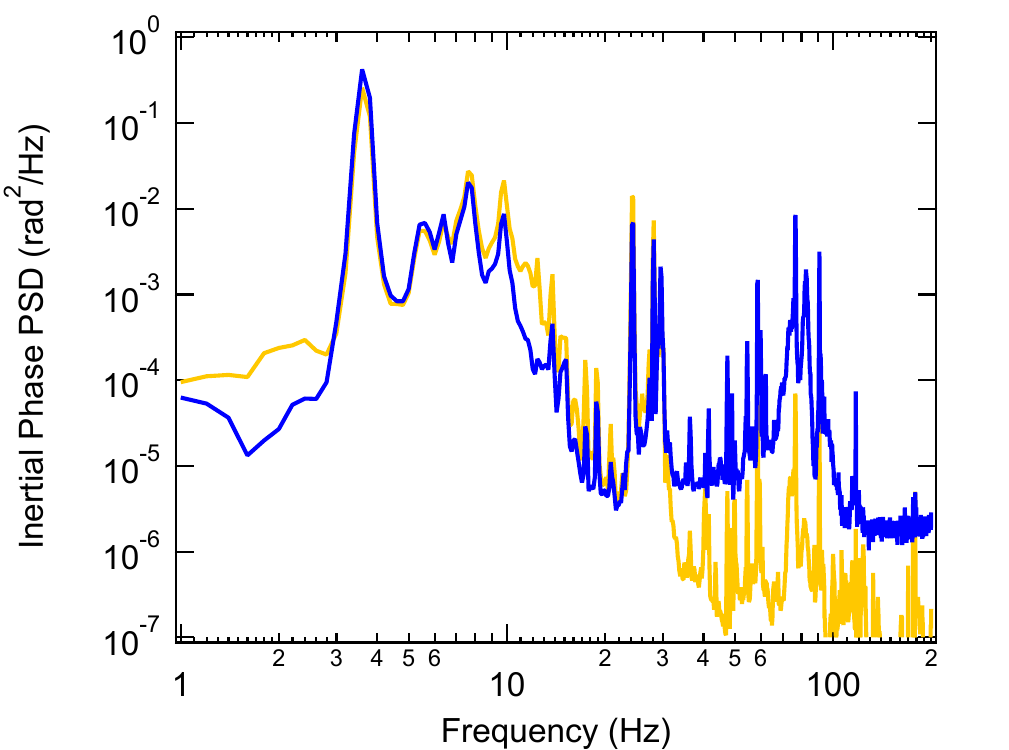}
\caption{\label{fig:inertWC} Power spectral densities are shown for the measured phase of the inertially-sensitive atom interferometer without k-reversal (blue) and for the phase prediction made using a pair of classical seismometers (yellow) mounted to a common surface below the atom interferometer. Rough agreement is seen from 3-30 Hz, with a general deviation above 30~Hz. Deviation at high frequency is attributed to differing mechanical modes of the seismometer mounting structures compared with the Raman mirrors and interferometer body itself. A minimum in the seismometer-derived atom interferometer phase response at $1/T=148~\mathrm{Hz}$ is described by Eq.~\ref{eq:phiseis}. The seismometer signals have been high-pass filtered at 2 Hz to avoid a strong low-frequency drift.
}
\end{figure}

An example plot of the power spectral density (PSD) of the seismometer-based phase estimation is shown as compared with the PSD of the inertially sensitive phase of the atom interferometer, acquired without k-reversal or added magnetic noise, in Fig.~\ref{fig:inertWC}. Many spectral features appear similarly in both the seismometer and atom interferometer. Between 3 and 15 Hz, motional modes of the sensor platform on passive vibration isolation feet (Newport VIBe Mechanical Vibration Isolators) appear. The height of these spectral features varies significantly from shot to shot, as reflected in both the seismometer and atomic spectra. Between 20 and 30 Hz, vibrational signals due to facility mechanical equipment including an air compressor appear consistently in both spectra. Under typical laboratory conditions, the standard deviation of the estimated inertial phase fluctuates between 100-200 mrad in the frequency range 1 Hz to 200 Hz. 

Various significant differences between the atom interferometer and seismometer-estimated phase spectra appear.  A notable dropoff of the seismometer-estimated spectrum occurs above 30~Hz, while the measured atomic phase spectrum exhibits a strongly peaked feature centered at 80~Hz. We hypothesize that this difference in the spectra is due to differences in inertial environment between the seismometer mounting points and the atom interferometer optomechanics. In this experiment, the Raman mirror was mounted outside of the vacuum chamber and various elements of the Raman beam optical path use adjustable optomechanics that are not highly rigid, potentially creating a response to the vibrational environment that is distinct from the seismometer response. Additionally, the seismometer-derived phase estimation shows a minimum in the PSD around the frequency $1/T$. This is a consequence of the phase calculation based on an assumption of rigid-body motion of the interferometer, Eq.~\ref{eq:phiseis}. While this is expected, the atom interferometer output spectrum does not exhibit such a strong minimum. This is, again, likely due to a violation of the assumption of rigid-body motion in the form of vibrations in the Raman optical paths, as well the presence of other technical noise sources such as laser intensity noise in the atomic inertial measurement. A new version of the atom interferometer apparatus, currently under development, moves key optics into the vacuum chamber,  increases rigidity of optomechanics, and incorporates optical power stabilization.

\section{Magnetic Field Noise}
\label{sec:magnetic}
In the presence of a time-varying magnetic field $\delta B(t)$ and spatially and temporally constant offset $B_0$, both aligned parallel to the Raman beams propagation direction, the atom interferometer phase shifts due to the quadratic Zeeman shift. The resulting interferometer phase is written simply as a function of the interferometer sensitivity function $g(t)$ \cite{cheinet_measurement_2008}:
\begin{equation}
\phi_b = \alpha \int_{-T}^T g(t) (B_0 + \delta B(t))^2 dt
\end{equation}
where, for convenience, we define $t=0$ as the time the atom passes through the central Raman beam.  $\alpha$ is the magnitude of the quadratic Zeeman clock shift, $\alpha = 2 \pi \times 575.15~\mathrm{Hz/G^2}$ in ${}^{87}$Rb.

Ignoring the finite duration of the Raman pulses,
\begin{equation}
g(t) = \begin{cases}  
-1, &-T \leq t < 0 \\
+1, &0 \leq t \leq T \\
0, &\text{otherwise}
\end{cases}
\end{equation}
where the sign reversal of the sensitivity function occurs due to the atomic population reversal induced by the central Raman $\pi$ pulse. The assumption of instantaneous Raman pulses is justified in the present work because all frequencies in the present analysis are much smaller than the inverse of the Raman pulse duration.

In the spatial-domain atom interferometer, the calculation of magnetically induced phase must account for the atomic velocity $v$ and the spatial distribution of the magnetic field. We define $\beta(x)$ to be the spatial profile of the magnetic field added by the noise coils that are centered on the final Raman beam, with $x$ being the coordinate along the approximately horizontal atomic trajectory. We model each coil as a thin current loop \cite{montgomery_useful_1961,smythe_static_1969} and normalize $\beta(x)$ so that $\delta B(t) = B_{coil}(t)\beta(v\times(t-T))$, where $B_{coil}(t)$ is the on-axis field measured at the center of the noise coil pair.

The magnetic-field-induced interferometer phase for an atom crossing the center Raman beam at time $t_0$ is then
\begin{equation}
\phi_b(t_0) = \alpha \int_{-T}^{T} g(t)\left[B_0 + B_{coil}(t+t_0) \beta (v\times(t-T))\right]^2 dt. \label{eq:phib}
\end{equation}
In practice, we calculate this phase using the discrete correlation of the measured temporally varying magnetic field with the calculated spatial profile and atom interferometer sensitivity function.

In the experiment, we use the noise coil pair described in Sec.~\ref{sec:apparatus} to apply a magnetic field with dc level 714 mG and ac standard deviation 48 mG with an approximately flat spectrum between dc and 300 Hz, as measured using a magnetometer (FW Bell 5180) placed at the geometric center of the noise coil pair. (We remove the noise coils from the vacuum chamber to measure magnetic field between the coils.) The applied magnetic field creates a dc phase shift of approximately $9.7~\mathrm{rad}$ and ac phase noise with standard deviation of 518 mrad as predicted by Eq.~\ref{eq:phib}. We use Eq.~\ref{eq:phib} to predict the power spectrum of the magnetically induced interferometer phase (Fig.~\ref{fig:waterfalls}~(d)) for the measured time-dependent magnetic field. The magnetically induced noise power is sufficient to almost entirely mask the atom interferometer's typical inertial signal spectrum in the laboratory - see Fig.~\ref{fig:waterfalls} (a). At frequencies above 100 Hz, the magnetic field noise appearing in the atom interferometer phase is attenuated compared with the magnetic field noise predicted based on magnetometer measurements. Potential sources for this discrepancy include the unmodeled magnetic permeability and inductance of the stainless steel vacuum chamber and copper gaskets used to seal vacuum flanges.

\section{Results}

\begin{figure*}
\includegraphics[width=.5\linewidth]{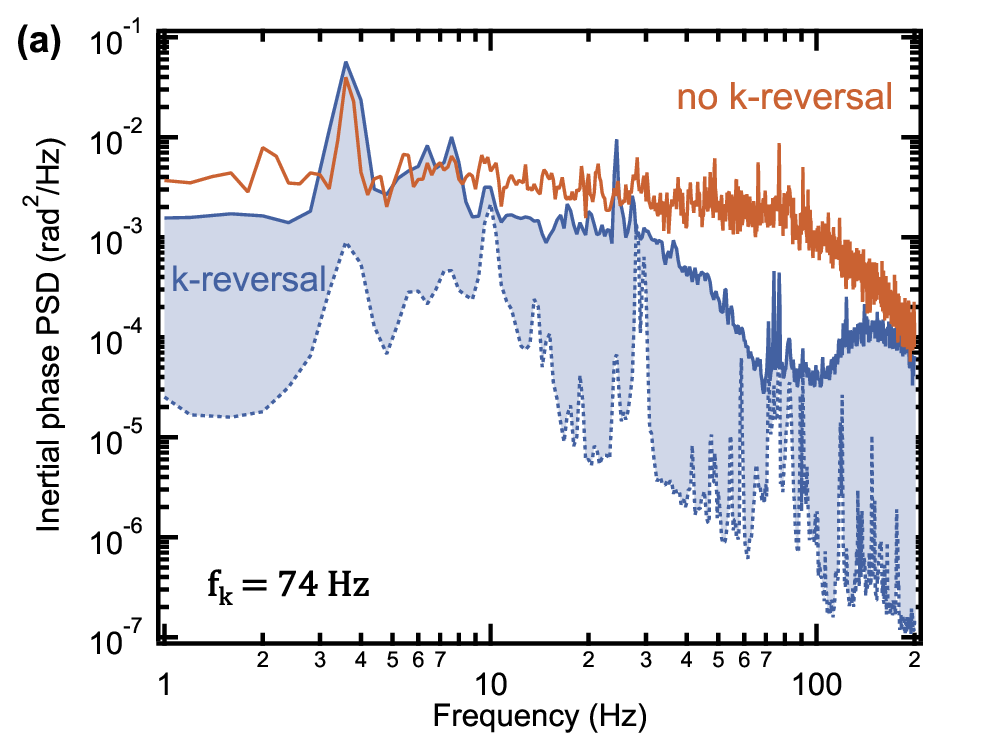}\includegraphics[width=.5\linewidth]{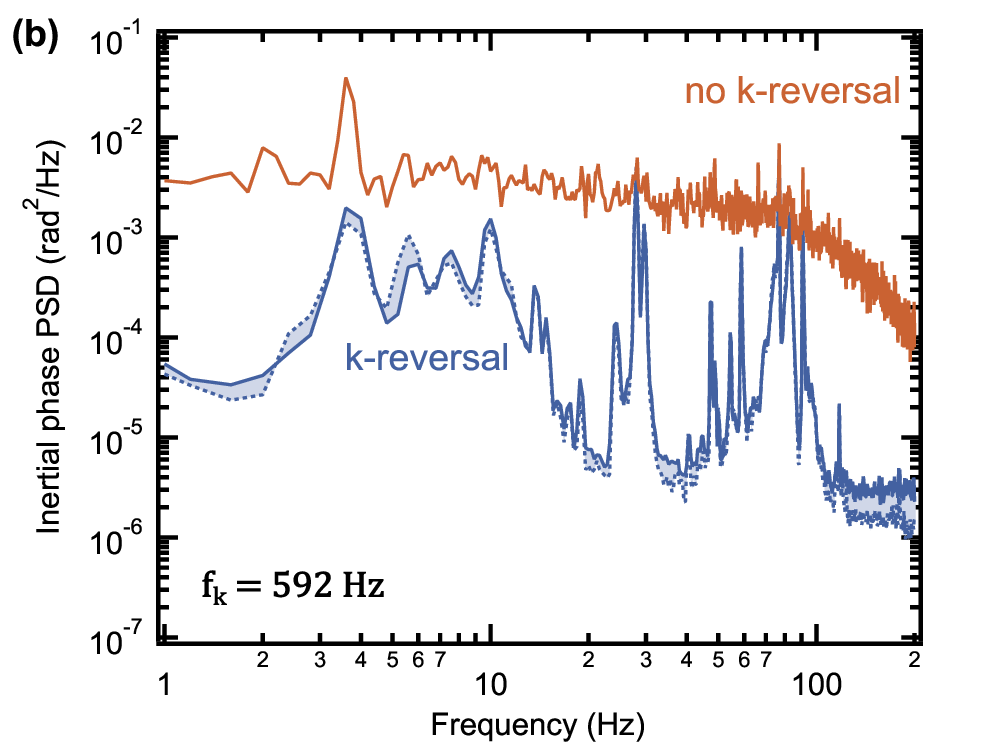}\\\includegraphics[width=.5\linewidth]{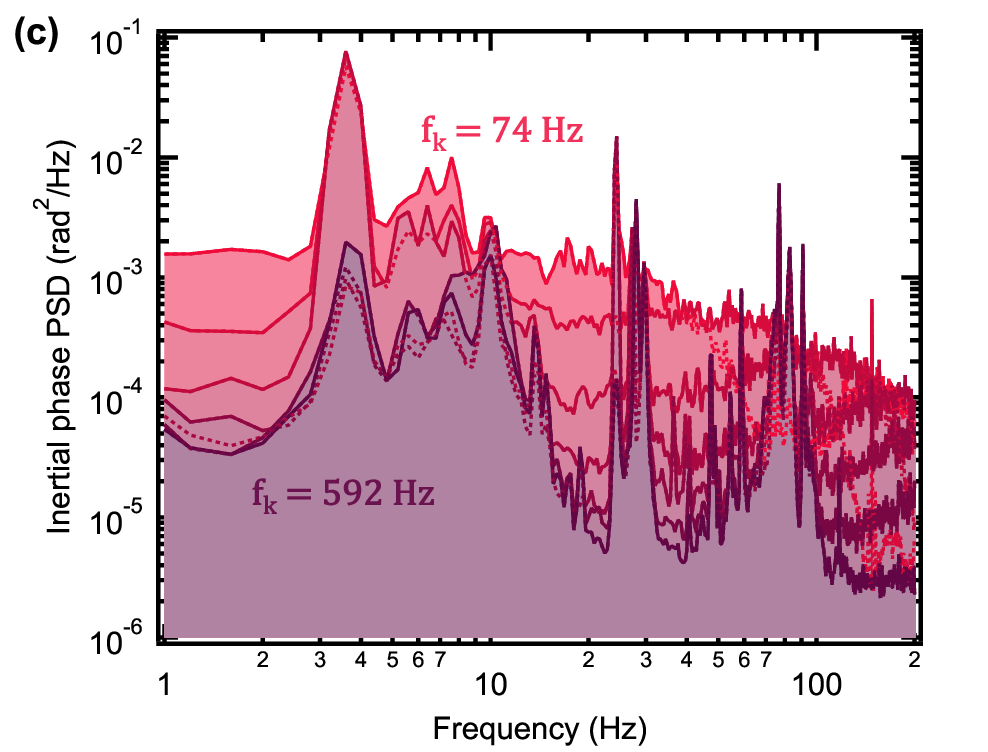}\includegraphics[width=.5\linewidth]{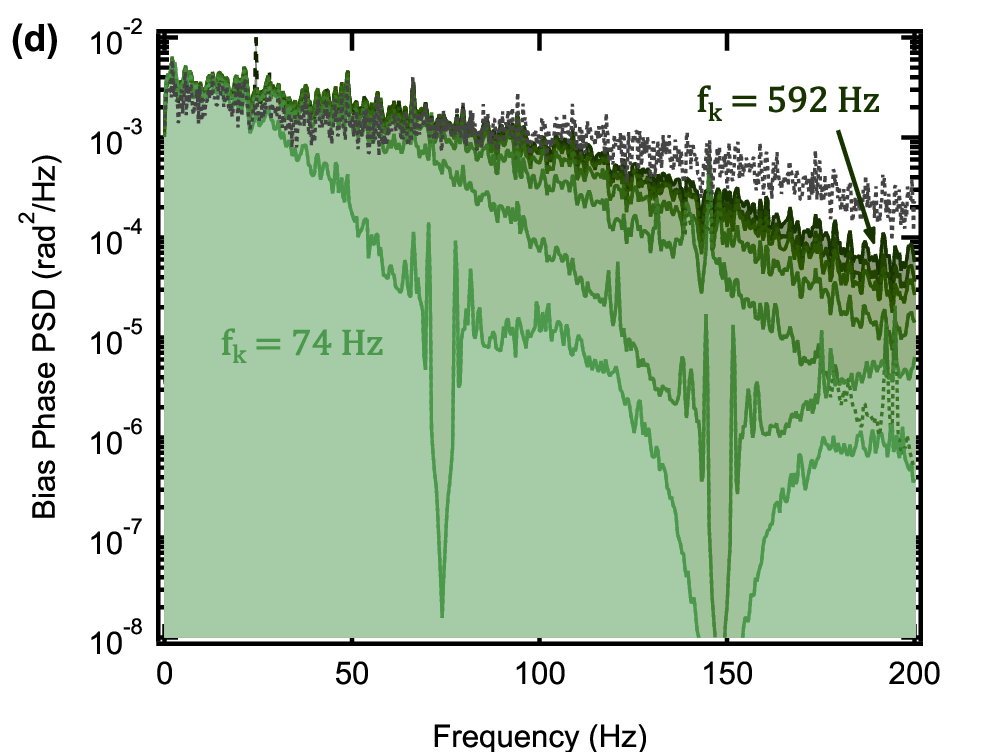}
\caption{\label{fig:waterfalls} The effect of k-reversal is demonstrated using the atom interferometer inertial phase PSD at k-reversal frequencies $f_k$ of 74 Hz (a) and 592 Hz (b). The inertial PSD with added magnetic noise is shown with (blue) and without (orange) k-reversal. For reference, the inertial signal is shown without added magnetic noise (dashed blue) and the area between the traces with and without added noise is filled in for clarity. PSDs of inertial phase (c) and bias phase (d) are shown in the presence of added magnetic field noise as a function of k-reversal frequency for multiples of 74 Hz up to 592 Hz. In each plot, the color gradient indicates k-reversal frequency, with darker colors corresponding to higher k-reversal frequency. The extremal reversal frequencies are labeled. The dashed gray line in (d) corresponds to the predicted bias phase PSD calculated from an independent measurement of the added magnetic field noise.
}
\end{figure*}

To evaluate the response of the sensor to the added magnetic noise and the laboratory's inertial noise environment, we plot the PSD of both the estimated inertial phase and estimated bias phase as a function of k-reversal frequency. We first illustrate the difference between slow k-reversal and fast k-reversal for two exemplary cases: $f_k=74$~Hz and $f_k=592$~Hz in Fig.~\ref{fig:waterfalls} (a) and (b) respectively, which display the PSDs of atomic inertial phase measured with and without added magnetic field noise, as well as the inertial phase measured with added magnetic noise but without k-reversal. With the noise coils turned on, the non-k-reversed inertial phase PSD is dominated by magnetic-field-induced phase noise, almost completely washing out the inertial signal. Both plots Fig.~\ref{fig:waterfalls} (a) and (b) show that k-reversal reduces the phase noise due to magnetic field across the displayed spectral range. However, the $f_k=74$~Hz k-reversal data show significantly more excess noise in the inertial signal than the $f_k=592$~Hz k-reversal data for in the added magnetic noise environment.

We display the atomic inertial and bias phase PSDs measured at the full range of k-reversal frequencies $f_k=n/2 T$ studied in this work in Fig.~\ref{fig:waterfalls} (c) and (d) respectively. In the bias phase spectrum, each successive increase in k-reversal frequency causes the estimated bias PSD to more closely approach the noise PSD predicted from the measured magnetic field time series. Roughly, the estimated bias phase PSD agrees with the predicted magnetic phase PSD for frequencies below $f_k/2$, the Nyquist-limited frequency for sampling the bias phase at rate $f_k$. Above $f_k/2$, significant attenuation of measured atomic bias phase occurs, as the bias phase estimation algorithm is not able to accurately measure the bias phase above the Nyquist limit. As discussed in Sec.~\ref{sec:magnetic}, even at the highest k-reversal frequency of 592~Hz, the bias phase PSD differs by a significant margin from the curve based on magnetometer measurements at frequencies over 100 Hz. However, this deviation is also evident in the phase measurement without k-reversal and is not attributable to the k-reversal algorithm. We also observe dips in the measured bias phase PSDs at frequency multiples of $f_k$. At these frequencies, the bias phase cannot be correctly estimated from the data as discussed in section \ref{Sec:kswitching}. For the PSDs with $f_k>200~\mathrm{Hz}$ in Fig. \ref{fig:waterfalls} (c) and (d), these resonances lie off-scale.

Inertial phase PSDs exhibit a different trend compared with the bias PSDs as a function of $f_k$, as shown in Fig.~\ref{fig:waterfalls}(c). The magnetic noise contribution to the inertial phase estimate is reduced monotonically as a function of k-reversal frequency, approaching a PSD similar to the noise-free atomic phase PSD at the highest k-reversal frequencies considered. A significant inertial noise reduction near dc is observed as $f_k$ is increased, as predicted by the error spectrum calculation plotted in Fig.~\ref{fig:alias}. This behavior contrasts with the bias phase PSDs in Fig.~\ref{fig:waterfalls} (d), in which the bias phase PSD most closely matches the predicted bias phase PSD for frequencies below $f_k/2$. For example, for $f_k=74~\mathrm{Hz}$, significant excess inertial phase noise exists for frequencies below $f_k/2$, even though the bias phase estimate PSD suggests that the noise spectrum is accurately captured in the bias phase estimate below $f_k/2$. Once again, this behavior is explained by the aliasing response curve of Fig.~\ref{fig:alias}, which indicates that bias phase at frequencies close to odd multiples of $f_k$ will be aliased down to near-dc frequencies in the inertial phase estimate. In this experiment with a 300 Hz bandwidth of additive magnetic field noise, all measurements with $f_k < 300~\mathrm{Hz}$ are expected to exhibit elevated noise levels near dc. This is broadly consistent with our observations.

Similarly to the bias phase PSDs, the inertial PSDs also display dips at $f_k$ as can be seen in Fig.~\ref{fig:waterfalls}. In the inertial PSD shown in Fig. \ref{fig:waterfalls} (a), we observe an overall reduction in inertially-sensitive noise density surrounding the k-reversal frequency (74 Hz in this case). This occurs because the signal processing algorithm, discussed in \ref{Sec:kswitching}, is based on the assumption that there is neither bias nor inertial phase content at the frequency $f_k$. Therefore, the dip in Fig. \ref{fig:waterfalls} (a) surrounding 74 Hz is due to a loss of information, not a result of any noise suppression. The measured inertial spectrum at $f_k=148$~Hz in Fig. \ref{fig:waterfalls} (c) likewise exhibits a corresponding dip in inertial response at 148~Hz.

\begin{figure}
\includegraphics[width=\linewidth]{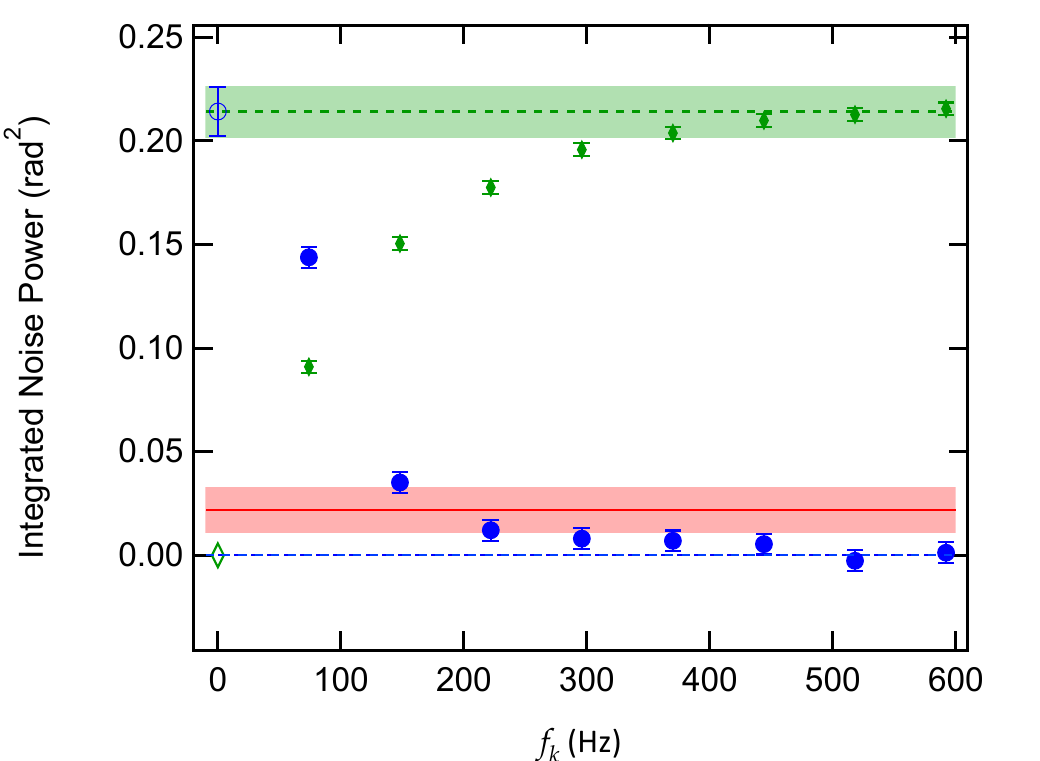}
\caption{\label{fig:integrated} Integrated power spectral density of atom interferometer output phases versus k-reversal frequency $f_k$, when magnetic field noise with 300 Hz bandwidth is applied. Integration is from 1 Hz to 100 Hz. Solid green diamonds: Integrated power spectrum of estimated atomic bias phase. Error bars are given by the standard deviation of the integrated spectrum of bias phase estimated from multiple 10-second intervals of magnetometer measurement. Solid blue dots: difference between integrated atomic inertial phase power spectrum with added magnetic field noise and without added magnetic field noise. Error bars are given by the standard deviation of the integral of the atomic inertial phase PSD, taken when the noise coil is turned off. Open diamond and dashed line at 0 $\mathrm{rad^2}$: indicates the value of integrated excess phase noise expected for ideal noise suppression. Open blue circle and dashed line at 0.214 $\mathrm{rad^2}$: difference between the integrated spectrum of inertial atomic phase in the presence of magnetic noise, measured without k-reversal, and the integrated spectrum of seismometer-estimated phase, measured simultaneously. Shaded green region and error bar on the open blue circle: standard deviation of the difference between integrated atomic inertial phase spectra and seismometer-calculated phase spectra, taken without added noise. Red solid line: Mean integral of the inertial phase spectrum estimated from seismometer measurements. Shaded pink region corresponds to $\pm1$ standard deviation of integrated seismometer phase spectra over several measurements.
}
\end{figure}

We quantify the suppression of magnetic field noise in the inertial phase estimate, as well as the inclusion of magnetic field noise in the estimation of bias phase, by considering the integrated PSD of interferometer phase over a frequency range of interest relevant to inertial measurements. Integrating the PSD from 1 Hz to 100~Hz captures a significant majority of the atom interferometer's inertial response. In Fig.~\ref{fig:integrated}, we observe the dependence of integrated noise power as a function of k-reversal frequency $f_k$. We plot the integral of the estimated bias phase PSD, as well as the integral of excess inertial phase noise - that is, the difference between the integrated PSD of inertial phase with and without added magnetic field noise. The noise captured as bias increases monotonically, becoming indistinguishable from the limit set by our estimate of the total amount of noise added into the system. This implies that, as $f_k$ increases, more of the added noise is estimated as bias phase $\phi_b$, suggesting improved removal of noise from the inertial phase estimate. Simultaneously, a monotonically decreasing trend is seen in the integrated excess inertial noise power, to the point where the observed excess inertial noise power is consistent with zero, and is smaller than the average integrated inertial phase power estimated from the classical seismometer outputs. For $f_k \geq 296~\mathrm{Hz}$, approximately the bandwidth of added magnetic field noise, further observed reduction in noise power becomes smaller than the error bars due to shot-to-shot fluctuations in the integrated inertial phase power. We attribute the shot-to-shot fluctuations in inertial phase to true changes in the inertial environment of the noisy laboratory, as they are exhibited by both the atomic inertial phase and the classical seismometer. The integrated bias noise power, on the other hand, continues increase until $f_k \geq 444~\text{Hz}$, suggesting that the true suppression of noise improves beyond the limit observable due to shot-to-shot inertial variations.

\section{Conclusion}

We have analyzed the effectiveness of atom interferometer k-reversal in the presence of time-varying inertial and bias phases, and we have demonstrated strong suppression of ac magnetic noise via rapid k-reversal of an atomic beam interferometer. While these results have focused on magnetic field noise due to the ease with which we can control magnetic field and predict magnetic-field-induced phase shifts, these results should be equally applicable to phase shifts due to differential ac Stark shifts and atomic collisions \cite{morinaga_sensitive_1993,weiss_precision_1994,peters_high-precision_2001}. In particular, we have demonstrated experimentally that k-reversal at high rate comparable to the noise bandwidth is advantageous for improving the resilience of the atom interferometer to noise. Further, noise content around odd integer multiples of the k-reversal frequency is aliased to near-dc frequencies, potentially inducing long-term errors. At the same time, increased k-reversal rate increases measurement dead time, which could introduce other aliasing noise sources or reduce signal-to-noise ratio. Further reductions in atomic temperature or alternative readout methods may ameliorate this concern.

In this work, the atomic sensor output is compared with signals from classical seismometers and a magnetometer, but without taking advantage of opportunities for sensor fusion between the atomic and classical sensors \cite{bidel_absolute_2018,cheiney_navigation-compatible_2018,tennstedt_integration_2021}. In the future, combining atomic sensor outputs with classical inertial sensors, as well as monitors of error sources such as laser intensity and magnetic field, has the potential to further reduce the effects of aliasing noise produced by k-reversal. It is also likely that modifications to the algorithm for bias phase estimation, such as alternative averaging window functions, may improve the aliasing behavior at the expense of measurement bandwidth. Finally, atom interferometry techniques that allow the simultaneous measurement of interference signals with both signs of photon recoil, or that do not change the internal state of the atoms, can eliminate challenges due to sequential k-reversal measurements, but often come at a cost in sensor complexity and measurement rate \cite{muller_atom-interferometry_2008,gebbe_twin-lattice_2020}.

\section*{Acknowledgments}
This work is supported in part by the Office of Naval Research and the Defense Innovation Unit. We thank Mark Bashkansky and Joshua Frechem for helpful comments.

%\bibliography{NRL5614_2}
%\bibliographystyle{apsrev4-2} %%For some reason 4-2 doesn't work on my computer. We can change before submission


\end{document}